\documentclass[12pt,preprint]{aastex}
\begin{document}
\newcommand{\msun}{M_{\odot}}
\newcommand{\kms}{\, {\rm km\, s}^{-1}}
\newcommand{\cm}{\, {\rm cm}}
\newcommand{\gm}{\, {\rm g}}
\newcommand{\erg}{\, {\rm erg}}
\newcommand{\kpc}{\, {\rm kpc}}
\newcommand{\mpc}{\, {\rm Mpc}}
\newcommand{\seg}{\, {\rm s}}
\newcommand{\kev}{\, {\rm keV}}
\newcommand{\hz}{\, {\rm Hz}}
\newcommand{\etal}{et al.\ }
\newcommand{\yr}{\, {\rm yr}}
\newcommand{\eq}{eq.\ }
\newcommand{\lya}{Ly$\alpha$\ }
\newcommand{\hi}{\mbox{H\,{\scriptsize I}\ }}
\newcommand{\hii}{\mbox{H\,{\scriptsize II}\ }}
\newcommand{\hei}{\mbox{He\,{\scriptsize I}\ }}
\newcommand{\heii}{\mbox{He\,{\scriptsize II}\ }}
\newcommand{\nhi}{N_{HI}}
\def\arcsec{''\hskip-3pt .}

\title{
On the evolution of the ionizing emissivity of galaxies and quasars required
by the hydrogen reionization}
\author{Jordi Miralda-Escud\'e}
\affil{The Ohio State University}
\email{jordi@astronomy.ohio-state.edu}

\begin{abstract}

  The average rate of emission of ionizing radiation per unit volume (or
emissivity) in the universe can be calculated as the ratio of the
intensity of the ionizing background to the mean free path of ionizing
photons. The intensity of the background is measured from the mean
transmitted flux of the \lya forest, and the mean free path is measured
from the abundance of Lyman limit systems, which has been observed so
far up to $z=4$. This yields an emissivity that is not larger than 7
ionizing photons per Hubble time for each atom in the universe at $z=4$,
which may reasonably arise from QSOs and star-forming galaxies. In order
for the reionization to end by $z=6$, and assuming that the clumping
factor of ionized gas during the reionization epoch is close to unity,
this ionizing comoving emissivity cannot decline from $z=4$ up to $z\sim
9$ by more than a factor $1.5$. If the clumping factor were much larger
than unity, then the emissivity would need to rapidly increase with
redshift. Unless the ionizing emissivity increases substantially from
$z=4$ to $z\sim 10 - 20$, the Thomson optical depth to the CMB must be
in the range $0.045 < \tau < 0.09$.

\end{abstract}

\keywords{cosmology: theory -- diffuse radiation -- intergalactic medium -- 
galaxies: formation}

\section{Introduction}

  The highest redshift quasar known at the present time, at $z=6.28$, is
the first one to show a complete Gunn-Peterson trough (Becker \etal
2001; Djorgovski \etal 2001). This suggests that the reionization of
hydrogen was probably completed near the epoch $z=6$. This conclusion is
not yet completely certain because an ionized medium can produce a
complete Gunn-Peterson trough if the neutral fraction is everywhere high
enough (owing to a low intensity of the ionizing background). However,
the abrupt change with redshift near $z=6$ of the fraction of
transmitted flux in the \lya region of the spectrum (Becker \etal 2001)
implies a rapid increase of the mean free path of ionizing photons with
time (Fan \etal 2002), at least in the region of the line of sight to
the $z=6.28$ QSO, just as expected when the last low-density regions of
the intergalactic medium are ionized (e.g., Gnedin 2000).

  Ending reionization by $z=6$ requires that at least one ionizing
photon has been emitted to the intergalactic medium for each baryon in
the universe, in addition to the amount needed to compensate for any
recombinations that have taken place. This implies that the emissivity
of ionizing photons (defined as the mean number of photons emitted per
unit time and volume) should be greater than the mean number density of
baryons divided by the age of the universe at $z \simeq 6$, unless the
emissivity had declined rapidly with time at $z > 6$.

  It is possible to infer the emissivity at lower redshift, by using the
intensity of the ionizing background measured from the mean transmitted
flux in the \lya forest, and the mean free path of ionizing photons
determined from the observed abundance of Lyman limit systems. The first
of these two quantities, the intensity of the ionizing background, can
be measured in the context of the present theoretical understanding of
the \lya forest as arising from a photoionized intergalactic medium that
follows the gravitational evolution of primordial fluctuations (see the
review by Rauch 1998 and references therein). The observed mean
transmitted flux can then be used to infer the \lya scattering optical
depth at the mean hydrogen density and in Hubble expansion, $\tau_u$:
\begin{equation}
\tau_u = 4.65\cdot 10^5 {\Omega_b h (1-Y)\, H_0\over H(z)}\, x_{HI} =
2.12 \left( {\Omega_b h^2 \over 0.02} \right)^2 {0.7\over h}
\left({0.3\over \Omega_m}\right)^{1\over 2}\, T_4^{-0.7}\,
\Gamma_{-12}^{-1} \left( {1+z \over 5} \right)^{9\over 2} .
\end{equation}
Here, $x_{HI}$ is the hydrogen neutral fraction at mean density, $T_4$
is the gas temperature in units of $10^4$ K, and $\Gamma_{-12}$ is the
photoionization rate due to the cosmic background in units of $10^{-12}
\seg^{-1}$. When the value of $\Omega_b h^2$ was still highly uncertain,
the value of $\tau_u$ inferred from observations of the mean transmitted
flux, $\tau_u\simeq 1.0 \pm 0.2$ at $z=3$ (see Rauch \etal 1997, and
Table 2 in McDonald \etal 2000) implied a lower limit to the value of
$\Omega_b h^2$ given by the lower limit on $\Gamma_{-12}$ that was
inferred from the observed QSOs. Now, $\Omega_b h^2$ has been more
accurately measured from the primordial deuterium abundance (O'Meara
\etal 2001) and the CMB spectrum of fluctuations (de Bernardis \etal
2002; Pryke \etal 2002; Spergel \etal 2003), and we can use the value of
$\tau_u(z)$ determined from the \lya forest to constrain the intensity
of the ionizing background and its evolution with redshift.

  The mean free path of ionizing photons is related to the abundance of
Lyman limit systems, which have column densities $N_{HI} > 1.6\times
10^{17}\cm^{-2}$. The reason is that the column density distribution of
absorption systems is approximately of the form $f(N_{HI})\, dN_{HI}
\propto N_{HI}^{-1.5}\, dN_{HI}$ (e.g., Petitjean \etal 1993), so the
average rate of absorption is dominated by high column densities, up to
the column density at which the absorption systems become optically
thick. The abundance of Lyman limit systems has been reasonably well
measured up to redshift $z=4$. In this paper, we combine the
measurements of the intensity and the mean free path of ionizing photons
to infer their rate of emission at $z=4$ (\S 2). We then compare this
emissivity to the observations of known sources at this redshift (\S 3).
Then, in \S 4, we find the consequences for the evolution of the
emissivity at higher redshift to satisfy the requirement of reionizing
the IGM by $z=6$. Finally, in \S 5 the consequences for the Thomson
optical depth of the Cosmic Microwave Background  photons are discussed.
We will use the flat cosmological model with $\Omega_m = 0.3$, $h=0.65$,
and $\Omega_b h^2 = 0.022$.

\section{The ionizing emissivity at $z=4$}

  In this section we evaluate the emissivity of ionizing radiation at
redshift $z=4$. In terms of the photoionization rate, $\Gamma$, and the
mean free path of the ionizing photons, $\lambda_i$, the photon
emissivity is given by $\epsilon_i = \Gamma/(\lambda_i \bar\sigma)$,
where $\bar\sigma$ is a frequency-averaged cross section. The result we
will find in this paper is that the emissivity at $z=4$ has a rather
small value, implying that it must not have declined much
with redshift, up to a substantially earlier epoch at $z>6$, in order to
satisfy the requirement that reionization was completed by $z\simeq 6$.
With that in mind, we will evaluate the maximum allowed value of
$\Gamma$ and of $\epsilon_i$ at $z=4$, which is the highest redshift at
which the mean free path $\lambda_i$ has been observationally
determined.

\subsection{The photoionization rate}

  The most recent measurements of the mean transmitted flux through the
\lya forest, $\bar F$, were reported by Bernardi \etal (2002), from a
large sample of QSOs of the Sloan Digital Sky Survey. From their Fig.\
4, the effective optical depth [defined as $\bar F \equiv
\exp(-\tau_{eff})$] at $z=4$ is $\tau_{eff} = 1.02 \pm 0.02$, where we
have estimated the error from the errorbars and scatter of their
measured points at various redshift bins. Hence, $\bar F < 0.37$ at
$z=4$. Note that previous values of $\bar F$ from a small sample of high
resolution spectra (Schaye \etal 2000, McDonald \etal 2000) were higher,
but this was probably because their method underestimated the QSO
continuum flux, which becomes increasingly difficult to set by fitting
regions between absorption lines as the redshift increases.

  We now use the results of McDonald \& Miralda-Escud\'e (2001,
hereafter MM01) to infer the photoinization rate $\Gamma_{-12}$.
According to their Figure 2, the value $\bar F < 0.37$ implies a
photoionization rate of $\Gamma_{-12}<0.40$ at $z=4.5$. This value
needs to be corrected from the model used in MM01 to the model we use
here, and to $z=4$, where we will choose all the parameters to maximize
$\Gamma_{-12}$, and therefore infer an upper limit. First of all, the
relationship between $\bar F$ and $\tau_u$ in equation (1) depends on
the amplitude of the mass power spectrum at the Jeans scale of the \lya
forest: a lower amplitude of fluctuations implies that the gas is less
concentrated in collapsed regions that correspond to saturated
absorption lines, and a greater fraction of the gas is left in
low-density regions, where it contributes more efficiently to decrease
the mean transmitted flux. The measured amplitude of the power spectrum
of the \lya forest (see Croft \etal 2002 and references therein) is
$0.86$ times lower than the amplitude in the model that was used by MM01
(as shown in McDonald \etal 2000), which had the parameters
$\sigma_8=0.79$ and $n = 0.95$ for the power spectrum. In other words,
the model used by MM01 would match the \lya forest power spectrum
amplitude found by Croft \etal if its normalization had been changed to
$\sigma_8=0.86 \cdot 0.79 = 0.68$, leaving all other parameters
unchanged. Since the amplitude of mass fluctuations grows proportionally
to the scale factor $a$ (the effects of the vacuum energy and the
radiation are negligible at $z=4$), the model with normalization
$\sigma_8=0.68$ has the same relation of $\bar F$ and $\tau_u$ at $z=4$
as the model with normalization $\sigma_8 = 0.79$ at $z = (1+4)/0.86 - 1
= 4.8$. It was also found in MM01 that the value of $\tau_u$ inferred
from a fixed $\bar F$ depends on the amplitude approximately as
$\sigma_8^{-2}$ (this can in fact be checked by comparing the values of
$\Gamma_{-12}$ at $z=4.5$ and $z=5.2$ for fixed $\bar F$ in Fig. 2 of
MM01). Because the limit $\Gamma_{-12} < 0.4$ was derived at $z=4.5$,
and $\Gamma_{-12}\propto \tau_u^{-1}$, the power spectrum amplitude
correction is a factor $(5.8/5.5)^2$. We note that here we have taken
the central measured value of Croft \etal of the power spectrum
normalization, instead of a lower value within their errorbar (which
would further raise the upper limit to $\Gamma_{-12}$ by about 20\%),
because other measurements of the power spectrum amplitude favor a
higher value instead (see Seljak 2001; van Waerbeke \etal 2001; Croft
\etal 2002; Jarvis \etal 2002; Spergel \etal 2003).

  Second, the inferred $\Gamma_{-12}$ depends on the gas temperature.
MM01 assumed $T_4=2$, but measurements of the gas temperature from the
\lya forest spectra have given values in the range $1.2 < T_4 <2$
(Theuns \etal 2002 and references therein). Although the optical depth
of a uniform medium is proportional to $T^{-0.7}$ (eq.\ 1) because of
the temperature dependence of the recombination coefficient, the
relation between $\bar F$ and $\tau_u$ is also affected by the
temperature because thermal broadening spreads some of the absorption
from saturated spectral regions to other regions with less absorption.
We use the result found in McDonald \etal (2001, their \S 7.4) that
thermal broadening causes a change in $\tau_u$ of 27\% of that due to
the recombination coefficient, in the opposite direction, therefore
giving an effective dependence $\Gamma_{-12} \propto T^{-0.5}$. If we
assume the lowest value of the temperature among the reported
measurements, $T_4=1.2$, then the temperature correction for the
upper limit of $\Gamma_{-12}$ is $(2/1.2)^{0.5}$.

  Finally, we correct for all other parameters used in MM01 ($\Omega_b
h^2 = 0.02$, $h=0.65$, $\Omega_m=0.4$, $z=4.5$) to the ones used here
($\Omega_b h^2 = 0.022$, $h=0.65$, $\Omega_m=0.3$, $z=4$), which we
choose within the current uncertainties to maximize the inferred
$\Gamma_{-12}$. The result of all these corrections (see eq.\ 1) for
the upper limit to $\Gamma_{-12}$ at $z=4$ is
\begin{equation}
\Gamma_{-12}(z=4) < 0.40 \left({5.8\over 5.5}\right)^2\, \left({2 \over
1.2}\right)^0.5\, \left({0.022 \over 0.02}\right)^2\, \left({0.4 \over
0.3}\right)^{1\over 2}\, \left({5 \over 5.5}\right)^{4.5} = 0.53 ~.
\end{equation}

\subsection{The mean free path}

  Storrie-Lombardi \etal (1994) found that the rate of Lyman limit
systems (defined as absorption systems with column density $\nhi > 1.6
\times 10^{17}\cm^{-2}$) at $z=4$ is $dN_{LL}/dz = 3.3\pm 0.6$. In other
words, the mean spacing between Lyman limit systems is $\lambda_{LL}= c
H^{-1}(z)/(1+z)/(3.3\pm 0.6)= (0.06\pm 0.01) c H^{-1}(z)$ at $z=4$.
To find the
relationship of $\lambda_{LL}$ to the mean free path of ionizing
photons, we assume that the column density distribution of the absorbers
is $f(\nhi)\, d\nhi \propto \nhi^{-1.5} \, d\nhi$ (Petitjean \etal
1993). The absorption probability per unit length, $\lambda_0^{-1}$, for
a photon at the hydrogen ionization edge frequency $\nu = \nu_{HI}$, is
\begin{equation}
{1 \over \lambda_0 } = {
\int_0^\infty d\tau \, \tau^{-1.5}\, \left( 1- e^{-\tau} \right) \over
\int_1^\infty d\tau \, \tau^{-1.5}\, \lambda_{LL} } =
{\sqrt{\pi} \over \lambda_{LL} } ~,
\label{mfp0}
\end{equation}
where $\tau = \nhi/(1.6\times 10^{17}\cm^{-2})$. At a higher frequency
$\nu$, and approximating the photoionization cross section as
proportional to $\nu^{-3}$, the mean free path is increased to
$\lambda_{\nu} \simeq \lambda_0 (\nu/\nu_{HI})^{1.5}$.

  We note here that our mean free path in (\ref{mfp0}) is 1.5 times
longer than that adopted in Madau, Haardt, \& Rees (1999), and in
Schirber \& Bullock (2002). For some reason, their opacity corresponds to
a Lyman limit system abundance of 5 per unit redshift, larger than
reported by Storrie-Lombardi \etal (1994).

\subsection{The emissivity}

  The photoionization rate is related to the emissivity
$\epsilon_{\nu}$, the cross section $\sigma_{\nu}$, and the mean free
path as a function of frequency by
\begin{equation}
\Gamma = \int_{\nu_{HI}}^\infty d\nu \, \epsilon_{\nu} \lambda_{\nu}
\sigma_{\nu} ~.
\end{equation}
Here, we use the approximation that the mean free path is much less than
the horizon distance (the exact relation between the emissivity and
background intensity involves an integration over time with redshift
effects, a full knowledge of the spectrum, and reemission; see
Miralda-Escud\'e \& Ostriker 1990; Haardt \& Madau 1996; Fardal, Giroux,
\& Shull 1998. We treat reemission approximately in \S 2.4). Assuming
that $\epsilon_{\nu} \propto \nu^{-\eta-1}$,
$\lambda_{\nu} \propto \nu^{1.5}$, and $\sigma_{\nu} \propto \nu^{-3}$, 
we have $\Gamma = \epsilon_0\lambda_0\sigma_0/(1.5+\eta)$,
where $\sigma_0=6.3\times 10^{-18}\cm^2$, and $\epsilon_0$ is the
emissivity per unit frequency at the ionization edge. The emissivity
integrated over frequency is $\epsilon_i = \int d\nu \epsilon_{\nu}=
\epsilon_0/\eta$. The emissivity can be conveniently
expressed as the number of ionizing photons emitted per Hubble time and
for each atom in the universe:
\begin{equation}
{\epsilon_i(z=4) \over H_4\, n_{at} } = { \sqrt{\pi}\, (\eta + 1.5 )\,
\Gamma \over (0.06\pm 0.01)\, \eta c \sigma_0 n_{at}\, (1+z)^3 } ~,
\end{equation}
where $n_{at}$ is the comoving number density of atoms (adding hydrogen
and helium), $n_{at} = \Omega_b \rho_{crit}/m_H (1-3Y/4)= 2.03\times
10^{-7}\cm^{-3}$, and $H_4$ is the Hubble constant at $z=4$. Using
$\Gamma_{-12} < 0.53$, taking the lower limit for the mean free path
(to obtain an upper limit for the emissivity), and assuming
$\eta = 1.5$, the final result is
\begin{equation}
{\epsilon_i(z=4) \over H_4 n_{at} } < 7.6 ~.
\label{emisi}
\end{equation}
In other words, there are fewer than eight ionizing photons being
emitted per Hubble time and for each atom at $z=4$.
  
\subsection{Effect of reemission from the \lya absorbers}

  Some of the photons that contribute to the photoionization rate in
the intergalactic medium arise from reemission by the \lya absorption
systems (see Haardt \& Madau 1996). For every hydrogen atom that is
photoionized by absorbing a photon, and assuming ionization
equilibrium, a recombination will take place, and about 40\% of the
recombinations are direct to the ground state, producing another
ionizing photon. However, the majority of the ionizing photons that
are not very close to the hydrogen ionization edge are absorbed deep
within Lyman limit systems, and the photons produced by recombinations
in these regions are close to the ionization edge [with a typical
frequency separation from the Lyman limit of $\Delta \nu /\nu \simeq
kT/(13.6\, {\rm eV}) \simeq 0.1$] and not likely to
escape from the Lyman limit system. Because of this, reemission by
hydrogen recombination increases the background intensity by only
$\sim 10\%$ (as is found, for example, by using the approximation that
the reemitted photons escape only when produced in an absorber of
column density $\nhi < 2/\sigma_0$, which is the column density for
which the typical optical depth for the photon to escape is unity).

  Some photons are also emitted by helium recombinations.
\hei makes little difference to the total budget of ionizing photons,
because for each absorption by \hei, one photon that ionizes hydrogen is
produced by the subsequent recombination. Because we did not include
absorption by \hei in our previous estimate of the mean free path, we do
not include the reemission either.

  Finally, photons with energy above $4$ Rydbergs are mostly used to
photoionize \heii, and each \heii recombination can produce between one
and three photons capable of ionizing hydrogen. About 25\% of the
recombinations that are not direct to the ground state go to $n=2$,
producing a Balmer continuum photon, which can ionize hydrogen. In
addition, all the recombinations produce either a Lyman series photon,
or two continuum photons when the last transition is from 2s to 1s; in
$\sim 8\%$ of the recombinations, two continuum photons that can both
ionize hydrogen are produced (see Osterbrock 1989, Tables 2.8, 4.11, and
4.12). Hence, on average 1.33 ionizing photons are produced for each
\heii that is photoionized. For a source spectrum $\epsilon_{\nu}\propto
\nu^{-\eta - 1}$, with $\eta=1.5$, the fraction of photons used to
ionize \heii is $1/8$, so \heii reemission increases the emissivity by
a factor $1+0.33/8=1.04$. This factor can be reduced during the \heii
reionization, when the rate of recombinations is less than the rate
of photoionizations.

  Based on this discussion of the effect of reemission from absorbers,
we reduce our upper limit in (\ref{emisi}) for the total emissivity
$\epsilon_i$ by a factor $1.14$, to obtain the upper limit for the
direct emissivity from sources $\epsilon_{si}$ at $z=4$ (rounded up
to one digit given the various uncertainties discussed),
\begin{equation}
{\epsilon_{si}(z=4) \over H_4 n_{at} } < 7 ~.
\label{emiss}
\end{equation}

\section{Comparison with the emissivity from
Lyman break galaxies and QSOs}

\subsection{QSOs}

  Fan \etal (2001) measured the following QSO luminosity function at
high redshift ($z \gtrsim 3.5$) from SDSS:
\begin{equation}
\psi_Q(L)\, dL = 7.8 \times 10^{-8} \mpc^{-3}
\left( {L \over L_{-26} } \right)^{-\gamma} \, 10^{-0.47(z-3) }
{dL \over L_{-26} } ~,
\label{lfun}
\end{equation}
where $L_{-26}$ is the luminosity of a QSO with absolute AB magnitude
$M_{1450}=-26$ at $\lambda=1450 \, {\rm \AA}$, and $\gamma=2.58$.
Using a power-law spectral index $\eta_Q=1.5$ for the QSO spectrum
at $\lambda < 1450 \, {\rm \AA}$ (Telfer \etal 2002 find
$\eta_Q = -1.57$ at $\lambda < 1200$ \AA for radio-quiet QSOs, and
the spectral slope is shallower at longer wavelengths; e.g., Schmidt,
Schneider, \& Gunn 1995), the total number of ionizing
photons (with $\lambda < 912\, {\rm \AA}$) emitted is $L_{-26} = 5.48
\times 10^{56} {\rm photons}/{\rm s}$.
The luminosity function in equation (\ref{lfun}) was observed down to
$L\simeq 0.5 L_{-26}$ by Fan \etal (2001); the total emissivity from QSOs
clearly depends on the luminosity at which the slope of $\psi_Q(L)$
becomes shallower than $L^{-2}$. If we assume the form $\psi_Q(L)
\propto [ (L/L_{Q*})^{\gamma-1} + (L/L_{Q*})^{\gamma} ]^{-1}$, which
produces a similar slope of $\psi(L)$ at low luminosities ($L < L_*$) as
observed at lower redshift (e.g., Boyle \etal 1988, 2000), then the
comoving emissivity is
\begin{equation}
\epsilon_{QSO} = 1.45\times 10^{-24}\, \left( {L_{-26} \over L_{Q*}}
\right)^{\gamma-2} \, 10^{-0.47(z-3)} \,
\int_0^\infty { x\, dx \over x^{\gamma-1} + x^\gamma } \,
{\rm photons}\seg^{-1}\cm^{-3} ~.
\end{equation}
Expressing this as the number of photons emitted per Hubble time and
per atom, at $z=4$, we find,
\begin{equation}
{ \epsilon_{QSO} \over H(z=4) n_{at} } = 0.61
\left( {L_{-26} \over L_*} \right)^{\gamma-2} ~.
\end{equation}
A lower limit to the emissivity from QSOs is obtained if $L_*\simeq
0.5 L_{-26}$, $\epsilon_{QSO}/(Hn_{at}) \gtrsim 1$. In order not to
exceed the upper limit obtained in the last section,
$ \epsilon / [H(z=4) n_{at} ] < 7$, we must require
$L_{Q*} > 0.015 L_{-26}$. We note, however, that low-luminosity QSOs may
be intrinsically self-absorbed at the Lyman limit, reducing their
contribution to the ionizing emissivity (see Alam \& Miralda-Escud\'e
2002).

  The number of faint AGNs has been constrained by Steidel \etal
(2002), who find that their abundance at $z=3$ and at an apparent AB
magnitude $R\simeq 23.9$ (or luminosities $L=0.018 L_{-26}$) is about
3\% of the Lyman break galaxies abundance at a magnitude $R=24.6$ at
$z=3$. Using the Lyman break galaxy luminosity function given in
the next subsection, we find that this implies that the turnover of the
QSO luminosity function should occur at $L_{Q*}=0.046 L_{-26}$, assuming
again the form $\psi_Q \propto
[ (L/L_{Q*})^{\gamma-1} + (L/L_{Q*})^{\gamma} ]^{-1}$. If the value
of $L_{Q*}$ were the same at $z=4$ as at $z=3$ (but with the QSO
abundance having declined as $10^{-0.47z}$ at all luminosities, as in
eq.\ [\ref{lfun}]), QSOs would contribute about half of the total
emissivity required at $z=4$ from equation (\ref{emiss}). There are
also limits from the Hubble Deep Field (Conti \etal 1999), but these
probe QSOs at much lower luminosities and do not provide a stronger
constraint on $L_{Q*}$. In summary, the available observations are
consistent with a contribution to the ionizing emissivity from QSOs
that does not exceed the upper limit in (\ref{emiss}).

\subsection{Lyman break galaxies}

  The galaxy ultraviolet luminosity function at high redshift has been
measured by Steidel \etal (1999). They fit the luminosity function to a
Schechter function,
\begin{equation}
\phi(L) \, dL = \phi_*\, \left( { L \over L_*} \right)^{-\alpha} \,
e^{-L/L_*} \, {dL \over L_* } ~.
\label{gflum}
\end{equation}
After converting to the flat cosmological model with $\Omega_m=0.3$, 
$h=0.65$, the parameters they find are $\phi_* = 8.6\times 10^{-4}
\mpc^{-3}$, $\alpha=1.6$, and $L_* = 1.4\times 10^{29} \erg\seg^{-1}
\hz^{-1}$ at $\lambda=1700\, {\rm \AA}$. To convert this to the ionizing
luminosity, we use the models of Bruzual \& Charlot (1993), for which
the luminosity per unit frequency decreases by a factor 10 from
$\lambda=1700\, {\rm \AA}$ to $912\, {\rm \AA }$, for an age of $10^9$
years (these models assume a Salpeter mass function up to a maximum mass
of $125 \msun$ and solar metallicity). Using also the approximation of
$L_{\nu} \propto \nu^{-2}$ at $\lambda < 912 \, {\rm \AA}$, we find the
ionizing luminosity is $L_* = 1.06\times 10^{54}\, {\rm photons}/\seg$.

  The form of the galaxy luminosity function (\ref{gflum}) is highly
uncertain at $L\lesssim L_*$, because the observations of Steidel
\etal (1999) are not deep enough to reach lower luminosities, and
galaxies down to $10^{-0.8}L_*$ are only detected in the Hubble Deep
Field (e.g., Madau \etal 1996). The Hubble Deep Field counts at $z=4$
are significantly below the prediction from the model in equation
(\ref{gflum}), but this may be due to galaxy clustering (Steidel \etal
1999; see their Fig.\ 8). If we include only galaxies with luminosities
larger than $(L_*, 10-{0.4} L_*, 10^{-0.8} L_*)$, the total emissivity
inferred from (\ref{gflum}) is
\begin{equation}
{ \epsilon_{gal} \over H(z=4) n_{at} } = (3.4, 8.5, 13.8) f_{esc}
~~~~~~~~~  [L > ( 1, 10^{-0.4}, 10^{-0.8}) L_* ] ~.
\end{equation}

  Steidel, Pettini, \& Adelberger (2001) found that Lyman break galaxies
may have escape fractions of ionizing photons as large as 50\%, from a
possible detection of Lyman continuum flux after coadding spectra of
several galaxies. If this result were true, the luminous galaxies with
$L > 10^{-0.8}L_*$ would already produce the maximum emissivity in
equation (\ref{emiss}). Alternatively, $f_{esc}$ may be much lower (see
Giallongo \etal 2002; note also that the value $f_{esc}\simeq 0.5$
measured by Steidel \etal 2001 was for the subset of bluest quartile of
the Lyman break galaxies, which may not be representative), and lower
luminosity objects may be responsible for the bulk of the emissivity.
More accurate determinations of the escape fraction, the abundance of
Lyman limit systems, and the inferred photoionization rate from the \lya
forest transmitted flux will be needed to resolve this question.

\section{Constraints on the emissivity evolution to end reionization at
$z=6$}

  The highest redshift QSO known at the present time (Becker \etal 2001)
shows a complete Gunn-Peterson trough at $z\simeq 6$, but the spectra of
all the sources at lower redshift have detectable transmitted flux
blueward of \lya, indicating that most of the volume in the
intergalactic medium had to be reionized by $z=6$. We now address the
contraint implied by this fact on the evolution of the ionizing
emissivity, in view of its value at $z=4$ (eq.\ \ref{emiss}).

\subsection{The reionization equation}

  Reionization proceeds by the creation of ionized regions around
individual sources (Arons \& Wingert 1972). The mean free path of
ionizing photons through the neutral medium is typically very short
compared to the distance between sources, so reionization proceeds by
the growth of the ionized regions and can be modeled in terms of the
fraction of the volume that is ionized at time $t$, $y(t)$, assuming
that the remaining fraction $1-y$ is completely neutral. Moreover,
because the emitted photons are used to ionize atoms almost immediately
after their emission, it follows that the change in the fraction of
atoms that are ionized in the universe is equal to the ionizing photon
emission rate per atom minus the average recombination rate of atoms.
If $\epsilon_a(t) \equiv \epsilon_i(t)/n_{at}$ is the comoving ionizing
photon emissivity per atomic nucleus, $R(t)$ is the average
recombination rate per atom in ionized regions (which includes the
effect of an effective clumping factor of the ionized gas), and $F_i$ is
the fraction of baryons that actually need to be ionized in the ionized
fraction $y$ of the universe (with the remaining $1-F_i$ fraction being
the baryons that are in dense systems, such as stars or dense gas in
galaxies, that do not need to be ionized by external sources), the
equation for the evolution of $y$ is (Madau, Haardt, \& Rees 1999;
Miralda-Escud\'e, Haehnelt, \& Rees 2000)
\begin{equation}
{d(yF_i) \over dt} = \epsilon_a - Ry ~.
\label{reieq}
\end{equation}
This equation (apart from the factor $F_i$) is analogous to the one
obtained for the problem of an expanding cosmological \hii region around
an ionizing source, which was discussed by Shapiro \& Giroux (1987) and
Donahue \& Shull (1987), where $y$ behaves as the volume of the \hii
region produced by a source of luminosity proportional to
$\epsilon_a(t)$. It is also identical to the equation of radiative
transfer, where $yF_i$ plays the role of the intensity, and $RF_i^{-1}$
and $\epsilon_a$ play the role of the absorption and emission
coefficients, respectively. The solution is found through the
substitutions $d\tau=R F_i^{-1}\, dt$, and $w=y F_i\, e^{\tau}$:
\begin{equation}
y(t)= F_i^{-1} \int_{t_i}^t dt'\, \epsilon_a(t')\, e^{\tau(t')-\tau(t)}
~,
\end{equation}
where $\tau(t)=\int_0^t\, dt'' F_i^{-1} R(t'')$, and $t_i$ is the
initial time when the first sources turn on. Assuming that any clumping
factor is constant, and that $\Omega_m(t)\simeq 1$ [e.g., for the
cosmological constant model with a present matter density
$\Omega_{m0}=0.3$, we have $\Omega_m(z=4) = 0.982$], the recombination
rate varies as $R(t) \propto t^{-2}$. Choosing $t_4\equiv t(z=4) =
(2/3) H_4^{-1}$ as a fiducial time, and $R_4=R(t_4)$, we have:
\begin{equation}
y(t)= F_i^{-1} \int_{t_i\over t_4}^{t\over t_4} d(t'/t_4)\,
{ 2\epsilon_a(t')\over 3H_4 } \,
\exp\left[ R_4 F_i^{-1} t_4 \left( {t_4\over t} - {t_4 \over t' }
\right) \right] ~.
\label{reisol}
\end{equation}

\subsection{The recombination rate}

  Before presenting solutions of the reionization equation, we discuss
the value that we adopt for the recombination rate, $R$. The average
recombination rate is affected by the clumping factor of the ionized
gas, which we define as: $C_l = <\! n_e^2 \! >_J/<\! n_e\! >_J^2$. The
subscript $J$ is used here to denote that the average is to be taken
only over those regions of space where matter is predominantly
photoionized by photons that have escaped to the intergalactic medium.
This restriction is necessary because if we did not impose it, the
clumping factor would be dominated by regions of dense gas that are
locally ionized by sources that were not included in the emissivity in
equation (\ref{emisi}), precisely because the radiation from these
sources is locally absorbed. In fact, in the absence of any restriction
on the regions over which one must average to obtain the clumping factor
of ionized matter, then the clumping factor would obviously be very
large and dominated by the densest stars. Once stars are eliminated from
consideration, there are still \hii regions that form in dense clouds in
the interstellar medium of galaxies around massive young stars, in which
all the emitted photons are immediately absorbed locally. This emission
is irrelevant for the reionization of the intergalactic medium, and it
is best not to include in equation (\ref{emisi}), and at the same time
not to include the dense gas ionized locally in the clumping factor.
This definition of the clumping factor inevitably involves an arbitrary
choice in defining when a photon is considered to be escaped, and
therefore when a region is predominantly ionized by escaped photons
rather than local ones. In practice, a photon can be defined to have
escaped once it reaches an intergalactic region with gas density below
some critical value. The calculation of the clumping factor can only
be obtained from full radiative transfer simulations of reionization,
but its value can differ among studies by large factors depending on
the definition that is adopted (see Gnedin 2000 and references therein).
If the restriction that escaped photons dominate the radiation intensity
is not imposed, then very large values of the clumping factor can be
obtained, which could only be used consistently by including the locally
absorbed photons in the emissivity.

  Here, we use the simple model of Miralda-Escud\'e \etal (2000) for
the clumping factor, which uses the approximation that the gas is
ionized only up to an overdensity $\Delta_i = \rho_i/\bar\rho$, and all
the gas at higher densities is neutral, or otherwise is ionized locally
and therefore does not contribute to the clumping factor. The
overdensity $\Delta_i$ is related to the mean free path $\lambda_i$ that
a photon can traverse before being absorbed. During reionization, the
mean free path $\lambda_i$ should be of order the size of the \hii
regions, or of order the typical separation between neighboring sources.
For $\lambda_i = 1000 \kms$, or $\sim 10 \mpc$ at $z=6$ in comoving units,
the critical overdensity is $\Delta_i\simeq 5$, and the clumping factor
is very close to unity (see Figs.\ 2d and 6 in Miralda-Escud\'e \etal
2000).

  It has been pointed out by Haiman, Abel, \& Madau (2001) that before
reionization is complete, a large number of low-mass halos should be
present and should contain dense gas accreted before it was heated by
reionization. The effect of these low-mass halos would be to decrease
the mean free path at a fixed overdensity $\Delta_i$, and to widen the
distribution of the overdensity. To estimate the plausible change in
the clumping factor due to these low-mass halos, we can assume that the
structure of the atomic intergalactic medium before reionization is
self-similar to that of the ionized intergalactic medium, but with the
scales reduced in proportion to the Jeans scale $\lambda_J$, so that the
clumping factor should depend only on the ratio $\lambda_i/\lambda_J$,
and the amplitude of the primordial density fluctuations at the Jeans
scale. If the temperature increases by a factor $100$ during
reionization (from $\sim 100$ K to $\sim 10^4$ K), the Jeans length
increases by a factor $10$.  The amplitude of primordial density
fluctuations in Cold Dark Matter with $\Omega_m=0.3$ and $h=0.65$
increases by a factor $\simeq 2$ between the scales of $0.05$ to $0.5$
Mpc (roughly corresponding to the Jeans scales at these two
temperatures). Therefore, the clumping factor of a medium in which all
the low-mass halos formed from scales intermediate between the Jeans
scales of the atomic and ionized medium have survived, at $z=7$ and
$\lambda_i=10^3 \kms$, should be the same as the clumping factor of a
medium with no such surviving halos, evaluated at $z=3$ and $\lambda_i
= 10^4 \kms$. From Figures 2b and 6 of Miralda-Escud\'e \etal (2000), we
find this clumping factor to be 7. We note that this is an upper limit
to the true clumping factor, because the gas in the low-mass halos that
might increase the clumping factor above the value of $\sim 1$ found
previously is likely to evaporate on a time short compared to the Hubble
time, once these halos have been reached by an ionization front (Shapiro
\& Raga 2001). The gas may also be expelled by a small amount of
internal star formation even before the halo is reached by external
ionizing radiation.

  Finally, a discussion is needed to decide if the case A or case B
recombination coefficient should be used. Ionizing photons of frequency
$\nu$ will typically be absorbed in regions of column density
$\nhi \sim 1.6\times 10^{17} \cm^{-2} (\nu/\nu_{HI})^3$, where
$\nu_{HI}$ is the Lyman limit frequency. The sources typically have
power-law spectra and most of the emitted photons have frequencies
substantially above $\nu_{HI}$, especially if the spectra are hardened
by internal absorption. Most of the photons emitted by direct
recombinations to the ground state, which have frequencies very close to
the Lyman limit, will therefore be produced in high column density
systems and will be reabsorbed within the same system (in addition,
among the small fraction of photons from direct recombinations to the
ground state that escape the system in which they have been produced,
some will be redshifted below the Lyman limit and will not be absorbed
again). Hence, these photons do not contribute to increase the cosmic
ionizing background intensity for the most part. An alternative way to
understand this argument is that, even though in principle we are in a
case B situation because the recombination photons are reabsorbed, in
practice the inclusion of the photons from recombinations increases the
clumping factor by the ratio of the case A to case B recombination
coefficients because these photons increase the ionization in dense
regions, therefore yielding a recombination rate equivalent to using
case A. Therefore, we use the case A recombination coefficient.

\subsection{Results for the reionization history}

  We now compute $y(t)$ assuming models where the emissivity varies
as $\epsilon_a \propto (1+z)^{-\beta} \propto t^{2\beta/3}$ for $z>4$,
and has the value $\epsilon_a/H_4 = 7$ at $z=4$, i.e., the maximum
allowed value found previously (eq.\ \ref{emiss}). We choose the
recombination coefficient $\alpha=3.15\times 10^{-13} \cm^3\seg^{-1}$,
valid for case A and $T=1.5\times 10^4$ K, and electron density $n_e =
6.95\times 10^{-5}\cm^{-3}$ at $z=6$, valid for $\Omega_b h^2 = 0.022$
and if helium is only once ionized. This yields a recombination rate
per Hubble time $R(z=6)/H(z=6) = 1.02$. In some cases, we increase
this recombination rate by a clumping factor $C_l$. We fix the fraction
of baryons in the ionized regions that actually need to be ionized to
$F_i=0.9$ (independent of redshift), in reasonable agreement with the
fraction of baryons that seem to be present in damped \lya systems at
$z\simeq 4$ (Lanzetta \etal 1991; Storrie-Lombardi \etal 1996), and the
fraction of baryons above the critical overdensity $\Delta_i\simeq 5$
at $z=6$ in the model discussed above for the clumping factor
(Miralda-Escud\'e \etal 2000, Fig.\ 2d).

  We solve equation (\ref{reisol}), starting at $y=0$ at some initial
redshift $z_i$ when the emissivity is assumed to turn on suddenly. The
parameters of the model that we vary are $z_i$, $\beta$, and $C_l$, and
we require all models to reach $y=1$ (i.e., the end of reionization) at
$z=6$.

\begin{figure}
\plotone{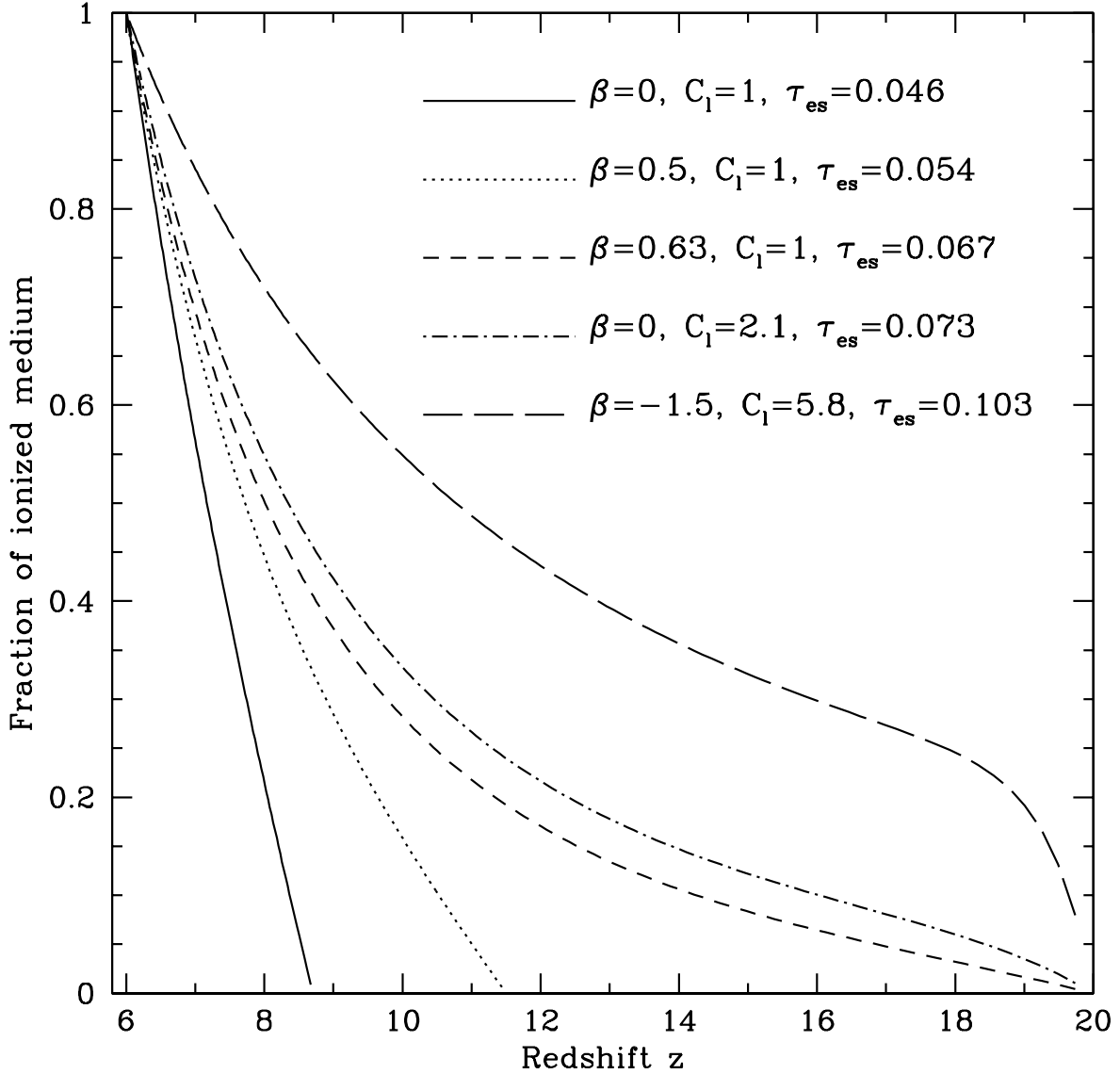}
\caption{Fraction of the volume of the universe filled by ionized regions
as a function of redshift. The models assume the emissivity is equal to
$\epsilon_a/H(t)= 7 [(1+z)/5)]^{-\beta}$, at $z<z_i$, where $z_i$ is the
initial redshift where the emission is turned on. The clumping factor of
the ionized gas is $C_l$.
The models are constrained to reach a fraction of unity at $z=6$. The
electron scattering optical depth to CMB photons is indicated for each model.
}
\end{figure}

  The results of five representative models are shown in Figure 1. When
$C_l=1$, assuming the emissivity to remain constant satisfies the
requirement of completing reionization by $z=6$ when the emissivity is
turned on at $z_i=8.7$. With only a slight decline of the emissivity,
the initial redshift needs to be pushed to much higher values. For
$\beta=0.63$, which corresponds to a decline of the comoving emissivity
by a factor of only $\sim 1.5$ between $z=4$ and $z=9$, the sources need
to start at very high redshift to reach $y=1$ at $z=6$. Naturally,
changes in the emissivity at redshifts higher than $\sim 15$ are
irrelevant for the requirement that reionization ends at $z=6$, because
most of the atoms ionized at this high redshift will recombine again.
We therefore conclude that, for the case of no clumping, {\it the
ionizing emissivity at $6 < z \lesssim 9$ cannot be much lower than its
maximum value allowed at $z=4$}. In other words, the sources responsible
for the emission of ionizing radiation cannot decrease their overall
comoving emissivity by more than a factor $\sim 1.5$ between $z=4$ and
$z=9$.

  Allowing now for a clumping factor, an increase to $C_l=2.1$ already
requires the model with constant emissivity ($\beta=0$) to maintain the
emissivity up to a very high initial redshift. A model where the same
amount of radiative energy is emitted at every Hubble time [i.e.,
$\beta=-1.5$, since $H(z)\propto (1+z)^{1.5}$], starting at very high
$z_i$, results in complete reionization at $z=6$ when $C_l=5.8$.
Therefore, for a clumping factor as large as $\sim 7$, as suggested
previously for a model where the ionized gas in all low-mass halos is
retained, the emissivity would need to substantially increase with
redshift.

\subsection{Comparison to the observed evolution at $z<4$}

  We have shown that by combining the measurement of the ionizing
emissivity $\epsilon_i$ at $z=4$ with the requirement that the
reionization must end by $z=6$, we can infer that the emissivity does
not decrease with redshift by more than a factor $\sim 1.5$ from $z=4$
to $z=9$. We can now ask how this evolutionary constraint compares to
the evolution at $z<4$ that is implied by observations of the \lya
forest decrement and the mean free path between Lyman limit systems.
From McDonald \& Miralda-Escud\'e (2001), the photoionization rate (or
proper intensity of the background) declines roughly as $(1+z)^{-1.5}$
from $z=2.5$ to $z=4$. The abundance of Lyman limit systems per unit
redshift is proportional to $(1+z)^{1.5}$ (Storrie-Lombardi \etal 1994),
implying that the proper mean free path varies as $(1+z)^{-4}$. Hence,
the comoving emissivity evolves as $(1+z)^{-0.5}$, essentially constant
within the likely errors of measurement for both the proper intensity
and mean free path evolution. Therefore, the results of \S 4.3 show that
if the clumping factor is low, this constant emissivity needs to roughly
maintained at the same value up to $z\sim 9$, while if the clumping
factor is much larger than one, the emissivity needs to modify its
evolution and increase strongly with redshift at $z>4$.

\subsection{The Thomson optical depth of the Cosmic Microwave Background}

  The optical depth of Cosmic Microwave Background photons to electron
scattering, $\tau_{es}$, is given by
\begin{equation}
\tau_{es} = \int_0^{z_i} dz\, {c\, dt\over dz} n_{e0} (1+z)^3
\sigma_{es}\, y(z)\, F_i ~,
\label{taues}
\end{equation}
where $\sigma_{es}$ is the Thomson cross section for electron
scattering. The values of $\tau_{es}$ for the five models in Figure 1
are indicated in the figure. We have assumed $F_i=0.9$ at all
redshifts; the 10\% of the matter not included is supposed to be
either gas that remains atomic or molecular in self-shielded regions,
or stars.

  These results show that the minimum allowed value of $\tau_{es}$,
which occurs for models where the emissivity turns on at a redshift
not much larger than 6, is $\sim 0.045$. The maximum values of the
optical depth depend on the emissivity at high redshift. If the
emissivity does not increase with redshift, then the ionized fraction
of the volume at high redshift needs to be small due to the fast rate
of recombinations (i.e., for constant emissivity, $y \propto
(1+z)^{-3}$, and the integral in eq.\ [\ref{taues}] converges rapidly
at high $z$). For a model with a constant emissivity up to $z_i=20$
and clumping factor equal to one, the optical depth can be increased
only up to $\tau_{es}= 0.09$ (in this model, reionization ends at
$z=7.3$, but the reionization end could always be delayed by increasing
the clumping factor as the fraction $y$ reaches a value close to unity).

  An emissivity increasing with redshift can obviously produce larger
optical depths, and as mentioned before the value of the emissivity at
$z\gtrsim 15$ does not affect the epoch at which reionization ends, so
in principle the optical depth could be arbitrarily high if enough
ionizing sources were present at very high redshift. However, in the
flat Cold Dark Matter model with $\Omega_m = 0.3$, which is strongly
supported by all observations of large-scale structure, the first stars
form only at $z\simeq 20$ from $3-\sigma$ peaks collapsing into the
first low-mass halos (e.g., Couchman \& Rees 1986; Bromm, Coppi, \&
Larson 1999; Abel, Bryan, \& Norman 2000; Ricotti, Gnedin, \& Shull
2002; Venkatessan, Tumlinson, \& Shull 2003), and so the emissivity
cannot continue to increase up to a very high redshift. At the same
time, observations indicate a cosmic star formation rate in galaxies
that starts to decline with redshift at $z\gtrsim 2$ (Madau, Pozzetti,
\& Dickinson 1998), although the ionizing emissivity might increase with
redshift even when the star formation rate decreases because of an
increasing escape fraction for ionizing photons.

  As this paper was being refereed, the results of the WMAP mission were
announced. Kogut \etal (2003) give a result $\tau_e=0.16\pm 0.04$ from a
model-independent analysis, although the result and the error can change
when fitted to different CDM models (Spergel \etal 2003). The value of
$\sim 0.08$ implied by a constant emissivity with clumping factor of one
up to very high redshift would represent a $2-\sigma$ deviation from the
value of Kogut \etal (2003). A value as high as $\tau_e=0.16$ would
clearly imply a substantial increase of the emissivity up to $z\sim 20$,
when only a small fraction of the mass has collapsed into halos that can
form stars in the CDM model.

\section{Conclusions}

  The observed mean transmitted flux of the \lya forest, when combined
with independently measured values of the mean baryon density, the
amplitude of the power spectrum, and the gas temperature, yields a
measurement of the intensity of the ionizing background. The abundance
of Lyman limit systems determines the mean free path of ionizing
photons. These two quantities together determine the ionizing
emissivity. We have argued in this paper that the emissivity measured in
this way at $z=4$ is at most 7 ionizing photons per atom per Hubble
time. We have then shown that the emissivity over the redshift range
$6<z \lesssim 9$ cannot be much lower than the value at $z=4$ in order
that the universe can be reionized by $z=6$, if the clumping factor of
ionized gas during the reionization epoch is close to unity. If the
clumping factor is larger than unity, then the ionizing emissivity
must increase with redshift between $z=4$ and $z\sim 9$.

  The electron scattering optical depth must be at least $0.045$, given
the presence of ionized intergalactic gas at $z<6$, and a reasonable
minimum redshift range over which sources ionize the universe at $z>6$.
If the emissivity of ionizing photons does not increase with redshift
at $z>4$, then the optical depth must be less than $0.09$. Measurements
of the optical depth from the CMB spectrum of temperature and
polarization fluctuations therefore provide a powerful diagnostic to
decide if the peak of the comoving ionizing emissivity occurred at
$z\sim 4$ (with a broad plateau extending up to $z\sim 9$ so that the
end of reionization can be reached by $z=6$), or at a much earlier
epoch. The recent WMAP result has an error that is still too large to
reach a definite conclusion, but if a value $\tau_e\simeq 0.16$ is
confirmed it implies that the comoving emissivity continues to rise
up to $z\simeq 20$.

\acknowledgements

  This work was supported in part by grant NSF-0098515.

\end{document}